\documentclass[conference,letterpaper]{IEEEtran}

\usepackage{datetime}

\usepackage[utf8]{inputenc} 
\usepackage[T1]{fontenc}
\usepackage{url}
\usepackage{ifthen}
\usepackage{cite}
\usepackage[cmex10]{amsmath} 
\usepackage{amsfonts}
\usepackage{amsmath, amsthm, amssymb}
\usepackage{graphicx}
\usepackage{enumitem} 
\usepackage{algorithm}
\usepackage[noend]{algpseudocode}

\usepackage{hyperref}
\hypersetup{pdftex,colorlinks=true,allcolors=blue} 
\usepackage{hypcap} 

\DeclareMathOperator*{\argmax}{arg\,max}

\newcommand\blfootnote[1]{%
	\begingroup
	\renewcommand\thefootnote{}\footnote{#1}%
	\addtocounter{footnote}{-1}%
	\endgroup
}

\begin{document}
\title{Anytime Decoding by Monte-Carlo Tree Search}

\author{%
  \IEEEauthorblockN{Aolin Xu}
  \IEEEauthorblockA{}
}

\maketitle

\begin{abstract}
An anytime decoding algorithm for tree codes using Monte-Carlo tree search is proposed.
The meaning of anytime decoding here is twofold: 1) the decoding algorithm is an anytime algorithm, whose decoding performance improves as more computational resource, measured by decoding time, is allowed, and 2) the proposed decoding algorithm can approximate the maximum-likelihood sequence decoding of tree codes, which has the anytime reliability when the code is properly designed. The above anytime properties are demonstrated through experiments.
The proposed method may be extended to the decoding of convolutional codes and block codes by Monte-Carlo trellis search, to enable smooth complexity-performance trade-offs in these decoding tasks.
Some other extensions and possible improvements are also discussed.
\end{abstract}

\blfootnote{xuaolin@gmail.com}
	
\section{Introduction}
\subsection{Tree code}
\begin{figure}[t]
	\centering
	\includegraphics[scale = 0.28]{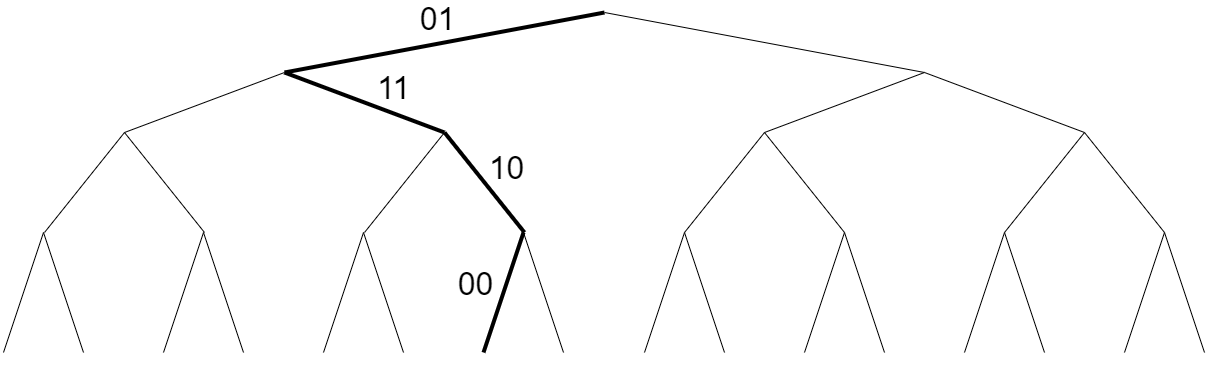}
	\caption{A tree code example with $k=1$, $n=2$, $d=4$. The encoding path for the information sequence $(0,1,1,0)$ is shown in bold.}
	\label{fig:tree}
\end{figure}

Tree code is a type of error correction code for which the encoding can be performed sequentially as tree traversal: starting from a root node, in the $i$th encoding step, the $i$th information symbol determines which child node to traverse, and the $i$th encoded symbol is taken as the label on the branch towards the chosen child node. For example, a rate $k/n$ tree code encoding $d$ information symbols defined over $\{0,1\}^k$ can be described by a regular $2^k$-ary tree of depth $d$, where each nonleaf node has $2^k$ child nodes and each branch has a label drawn from $\{0,1\}^n$.
Figure~\ref{fig:tree} shows such an example with $k=1$, $n=2$ and $d=4$.
Many familiar error correction codes can be represented as tree codes. Examples include the convolutional codes \cite{Wozencraft1957Sequential} and the block codes, both linear \cite{BCJR} and nonlinear \cite{trellis_blk}, as both types of codes have trellis representations, which can be expanded to trees.

An interesting property a tree code can have is the \emph{anytime reliability} \cite{Sahai_Mitter_06}, meaning that the probability of decoding error of each information symbol can be exponentially smaller as the delay for decoding that symbol linearly increases.
Tree codes with anytime reliability are shown to be essential for interactive communication over noisy channels \cite{Schulman96_iter_comm} and for stabilizing an unstable linear system over a noisy channel \cite{Sahai_Mitter_06}.
The anytime reliability of random convolutional codes over binary symmetric channel is claimed in the original paper where the convolutional code was proposed \cite{Elias_conv}. The existence of tree codes with explicit distance properties that guarantee anytime reliability under maximum-likelihood sequence decoding is proved in \cite{Schulman96_iter_comm}. Recent studies on more explicit constructions of codes having such a property include \cite{LTI_at_codes16,at_ldpccc,at_spcc}.

\subsection{Decoding by tree search}\label{sec:intro_MLSD}
A natural choice for decoding a tree code is the maximum-likelihood sequence decoding (MLSD). 
Over a discrete memoryless channel (DMC), the MLSD reduces to the minimum distance decoding, which is equivalent to searching for the minimum-length path from the root node to a leaf node. The length of each candidate path is the summation of the Hamming distance between the label on the $i$ branch in that path and the $i$th received symbol, over $i=1,\ldots,d$.
The complexity of full search on a $2^k$-ary tree of depth $d$ is proportional to the number of leaves, which is $O(2^{kd})$.

To manage the decoding complexity at the expense of compromising the optimality in the sense of MLSD, a number of approximate tree-search decoding algorithms have been proposed, mostly in the early literature of convolutional code decoding \cite{ECC_book_ShuLin} \cite[Chapter~8]{digit_comm_book5}. Prominent examples are sequential decoding and stack decoding, the complexity of which are random variables and are likely to be small when the channel noise is light. Other examples include feedback decoding and majority logic decoding or threshold decoding, where the complexity can be explicitly controlled by restricting the search depth, which however results in inferior decoding performance.
This paper presents a novel approximate tree-search decoding method, for which the complexity, determined by the run time, can be explicitly controlled, and the performance constantly improves as the run time increases. The method is based on Monte-Carlo tree search.

\subsection{Monte-Carlo tree search}
Monte-Carlo tree search (MCTS) is a heuristic tree search algorithm first proposed in the filed of computer game playing \cite{Remi_mcts06} \cite{Kocsis06banditbased}. It served as a major workhorse for the recent breakthroughs of computer Go \cite{AlphaGo} \cite{AlphaGoZero}.
In MCTS, each node in the tree represents a state in a game or a decision-making problem, while each branch emanating from that node corresponds to a move or an action, and is associated with a reward. The goal is to find a move from the root node with the largest expected accumulated reward towards a leaf node.
The core ideas behind MCTS are selecting the most promising move at each node while maintaining sufficient exploration during the search, and expanding the search tree based on random samplings of the search space. 
The major advantage of MCTS is its low complexity. For a $2^k$-ary tree of depth $d$, the complexity of MCTS is $O(md)$, where $m$ is the total number of rounds of search to run.
This is in contrast to the exponential complexity of $O(2^{kd})$ for the full tree search.

At the same time, MCTS is an \emph{anytime algorithm}, meaning that it can be stopped at any time and would return the best result it has found so far, and its result gradually improves if more rounds of search are run.
Being an anytime algorithm is a desirable feature for the algorithms used in intelligent systems \cite{Zilberstein_1996}, as it allows for flexible trade-offs between complexity and performance of an algorithm.

\section{Decoding by Monte-Carlo tree search}
\subsection{Tree search as sequential decision making}
As reviewed in Section~\ref{sec:intro_MLSD}, for a tree code transmitted over a DMC, the MLSD is equivalent to the minimum-length path search.
Searching for a minimum-length path over a tree can be further cast as a sequential decision-making problem as follows.
Each node $s$ in the tree represents a state in this decision-making problem; the branches emanating from this node towards its child nodes represent the available actions $\mathcal A(s)$ at this state; and each branch $a\in\mathcal A(s)$ is associated with a reward 
\begin{align}\label{eq:rwd}
r(s,a) = n - d_{\rm H}(x(s,a),{\bf y}_i) ,
\end{align}
where $d_{\rm H}$ denotes the Hamming distance, $x(s,a)$ is the label on the branch $a$, and ${\bf y}_i$ is the $i$th received encoded symbol, assuming that the branch $a$ is at depth $i$ in the tree.
At each state, once an action is chosen, the next state becomes the child node where the chosen branch leads to.
The goal is to sequentially choose actions from the root node to a leaf node to maximize the accumulated reward.

This sequential decision-making problem can be straightforwardly solved by dynamic programming through backward tracking.
Define $V^*(s) = 0$ for all the leaf nodes $s$.
For each nonleaf node $s$ and each $a\in\mathcal A(s)$, define recursively
\begin{align}\label{eq:Q*_def}
Q^*(s,a) = r(s,a) + V^*(s') ,
\end{align}
where $s'$ is the next state when $a$ is chosen at state $s$, and 
\begin{align}
	V^*(s) = \max_{a\in\mathcal A(s)} Q^*(s,a) .
\end{align}
With the function $Q^*$, the optimal actions of the sequential decision-making problem can be found as
\begin{align}
\hat a^*_{i} = \argmax_{a\in\mathcal A(s)} Q^*(s,a)
\end{align}
for $i=1,\ldots,d$, where $s$ evolves from the root node to a leaf node according to each $\hat a^*_i$.
The sequence $(\hat a^*_1, \ldots, \hat a^*_d)$ form an MLSD solution.

On a $2^{k}$-ary tree with depth $d$, the complexity of the above computation is $O(2^{k(d+1)})$, as there are $2^{kd}-1$ nonleaf nodes to evaluate the $Q^*$ function, and for each node there are $2^k$ entries.
This decoding method will serve as the benchmark for both the complexity and the performance in the experiments.

\subsection{Proposed decoding algorithm}
\subsubsection{Single-round decoding}
We first consider a basic decoding algorithm based on MCTS, the single-round decoding, which is an approximation of MLSD.
It is adapted from the \emph{upper confidence bound for trees} (UCT) implementation of MCTS \cite{Kocsis06banditbased}.
The algorithm involves running many rounds of search, during which an estimate of the $Q^*$ function defined in \eqref{eq:Q*_def}, denoted as $Q(s,a)$, is constantly updated.
Each round of search starts from the root node and ends at a leaf node, and can be executed in a recursive manner.
There are three stages in each round of search:
\begin{itemize}[leftmargin=*]
	\item
	\emph{Selection.}
	If the current state in the search is in the set $\mathcal T$, initially empty, then the search enters the selection stage.
	Otherwise it proceeds to the expansion stage. 
	During the selection stage, the function $Q(s,a)$ is updated for the states and actions visited and tried in the search. 
	The number of times $N(s,a)$ an action $a$ has been taken at a state $s$ is also tracked.
	During the selection, at each state, the action that maximizes 
	\begin{align}\label{eq:action_select}
		Q(s,a) + C \sqrt{\frac{\log N(s)}{N(s,a)}} ,
	\end{align}
	is taken, where $N(s) = \sum_{a\in\mathcal A} N(s,a)$ is the total number of times the state $s$ has been visited, and $C$ is a parameter that controls the amount of exploration in the search.
	The second term in \eqref{eq:action_select} can be thought as an {exploration bonus} that encourages selecting actions that have not been tried as frequently.
	Once an action is taken, a reward as in \eqref{eq:rwd} is received, and the state transits to the child node where the action leads to.
	The selection stage can be executed recursively until a leaf node or a node not in $\mathcal T$ is reached \cite{adp_samp_mdp,dmuu05}.
	At the end of each selection stage, the accumulated reward is returned either from a sub-selection stage or from the exploration stage, and is used to update the value for $Q(s,a)$.			
	
	\smallskip
	\item
	\emph{Expansion.}
	Once a state $s$ that is not in the set $\mathcal T$ is reached, the function values $N(s,a)$ and $Q(s,a)$ are initialized with $N_0(s,a)$ and $Q_0(s,a)$, respectively, by iterating over all of the actions available at that state.
	The initializing functions $N_0$ and $Q_0$ can be simply all-zero, or can be based on any available prior knowledge of the decoding problem.
	The state is then added to the set $\mathcal T$.
	
	\smallskip
	\item
	\emph{Exploration.}
	After the expansion stage, the actions are selected according to some default exploration policy $\pi_0$.
	The exploration policy does not have to be close to optimal, but it is a way to bias the search into areas that are promising.
	It can be stochastic, e.g.\, sampling $a$ from $\mathcal{A}$ uniformly at random, or can be in a deterministic round-robin fashion.
	The reward generation and state transition in the exploration stage is the same as in the selection stage.
	The exploration stage can also be executed recursively, until a leaf node or a desired depth is reached.
	
\end{itemize}

The single-round decoding is run until some stopping criterion is met, which could simply be a fixed number of rounds of search. 
The information symbols can then be decoded sequentially by taking the action that maximizes $Q(s,a)$ at each state, from the root node to an leaf node.

Algorithm~\ref{alg:mcts} below presents a detailed example of the single-round decoding method described above, where the number of rounds of search is fixed to $m$. The tree is assumed to have depth $d$, and the received sequence of encoded symbols is denoted by $\bf y$.

\begin{algorithm}
	\caption{Single-round decoding by MCTS}\label{alg:mcts}
	\begin{algorithmic}[1]
		\State\textbf{global variables}: $\mathcal T, N(\cdot,\cdot), Q(\cdot,\cdot)$
		
		\Function{DecodeByMCTS}{$d$, $m$, $\bf y$}
		
		
		\State $i \gets 0$, $\mathcal T\gets \emptyset$, $N(\cdot,\cdot) \gets 0$,  $Q(\cdot,\cdot)\gets 0$
		
		
		\While{$i < m$}
		\State \Call{Search}{root, $d$, $\bf y$}
		\State $i \gets i+1$
		\EndWhile\label{mctswhile}
		
		
		\State $s_1 \gets {\rm root}$
		\For{$j=1$ to $d$}
		\State $\hat a_j \gets \argmax_{a\in\mathcal A(s_j)} Q(s_j,a)$ 
		\State $s_{j+1} \gets$ \Call{NextState}{$s_j$, $\hat a_j$}
		\EndFor
		\State\Return $(\hat a_1, \ldots, \hat a_d)$
		\EndFunction
		
		\medskip
		
		\Function{Search}{$s$, $d$, $\bf y$}
		\If{$d=0$}
		\State \Return $0$
		\EndIf
		\If{$s\notin\mathcal T$}
		\For{$a\in\mathcal A(s)$}
		\State $(N(s,a), Q(s,a)) \gets (N_0(s,a), Q_0(s,a))$
		\EndFor
		\State $\mathcal T \gets \mathcal T \cup \{s\}$
		\State\Return \Call{Explore}{$s$, $d$, $\bf y$}
		\EndIf
		\State $N(s) \gets \sum_{a\in\mathcal A(s)} N(s,a)$
		\State $a \gets \argmax_{a\in\mathcal A(s)} Q(s,a) + C\sqrt{\frac{\log N(s)}{N(s,a)}}$ 
		\State $r \gets $ \Call{Reward}{$s,a, \bf y$}
		\State $s' \gets $ \Call{NextState}{$s,a$}
		\State $q \gets r  \,\, + $ \Call{Search}{$s'$, $d-1$, $\bf y$}
		\State $N(s,a) \gets N(s,a) + 1$
		\State $Q(s,a) \gets (1-\frac{1}{N(s,a)})Q(s,a) + \frac{q}{N(s,a)}$
		\State\Return $q$
		\EndFunction\label{fn:sim}
		
		\medskip
		
		\Function{Explore}{$s$, $d$, $\bf y$}
		\If{$d=0$}
		\State\Return $0$
		\EndIf
		\State $a\sim \pi_0(s)$
		\State $r \gets $ \Call{Reward}{$s,a, \bf y$}
		\State $s' \gets $ \Call{NextState}{$s,a$}
		\State\Return $r \,\, + $ \Call{Explore}{$s'$, $d-1$, $\bf y$}
		\EndFunction
		
	\end{algorithmic}
\end{algorithm}

\subsection{Sliding-root MCTS decoding}
In the single-round MCTS decoding described above, each search starts from the root node and ends at a leaf node, and finally all $d$ information symbols are decoded.
When $d$ is large, this method may not work well with an affordable number of rounds of search; especially for the information symbols with large indices, i.e. those near the leaf nodes, the decoding error can be high.
One way to improve is performing multiple rounds of MCTS decoding, where the root of the search in the successive rounds of decoding slides toward the leaf nodes according to the previously decoded information symbols; moreover, the depth of search can be smaller than $d$, and different for different rounds of decoding.

For example, a total number of $d$ rounds of decoding can be performed, where in the $i$th round of decoding, each round of search starts from a node in the $i$th level of the tree, determined by the previously decoded information symbols, and in the end of the $i$th round of decoding only the $i$th information symbol is decoded. The depth of search in the $i$th round of decoding can be $d+1-i$, which allows it to search to a leaf node, or it can be some number smaller than $d+1-i$.

\subsection{Support for anytime reliability}
To support anytime reliability, the decoding does not need to wait for receiving all $d$ encoded symbols to start. Instead, the $i$th information symbol can start to be decoded immediately after the $i$th encoded symbol is received.

One specific way to support anytime reliability is to perform $d$ rounds of MCTS decoding, where the $i$th round of decoding starts immediately after the $i$th encoded symbol is received, and each search in that round starts from the root node and ends at a leaf node in the $i$th level in the tree, namely, the depth of search is $i$. In the end of the $i$th round of decoding, the first $i$ information symbols are decoded. The larger the $i$ is, the more reliable the decoded information symbols are. Moreover, the earlier an information symbol is sent, the more reliably it can be decoded.

\section{Experiments}
Two tree codes are used for performing experiments on the proposed algorithm. Both of them are binary codes with rate $1/2$, namely $k=1$ and $n=2$. One of them has depth $10$, and the other has depth $25$. Each code is selected from a pool of randomly generated codes with the specified parameters to have the best decoding performance under MLSD.
The codeword is sent over a BSC with crossover probability $0.1$. For MCTS decoding, the parameter $C$ in \eqref{eq:action_select} is set to be in the same order as the search depth.

\begin{figure}[h]
	\centering
	\includegraphics[scale = 0.62]{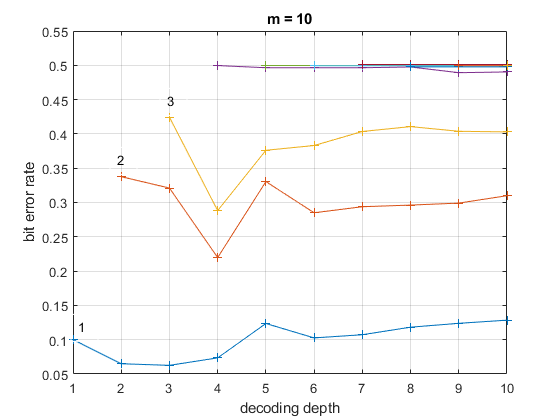}
	\caption{BER for multiple rounds of MCTS decoding supporting anytime reliability, with $d=10$ and $m=10$.}
	\label{fig:d10_6_c10_p1_m10_n120000_hi}
\end{figure}

\begin{figure}[h]
	\centering
	\includegraphics[scale = 0.62]{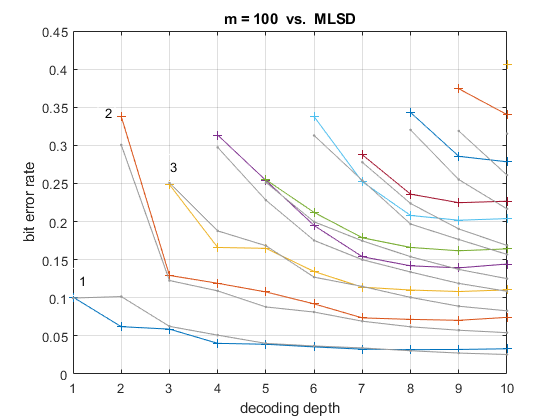}
	\caption{BER for multiple rounds of MCTS decoding supporting anytime reliability, with $d=10$ and $m=100$. Grey curves indicate BER for MLSD.}
	\label{fig:d10_6_c10_p1_m100_n120000_hi_vs_ml}
\end{figure}

\begin{figure}[h]
	\centering
	\includegraphics[scale = 0.62]{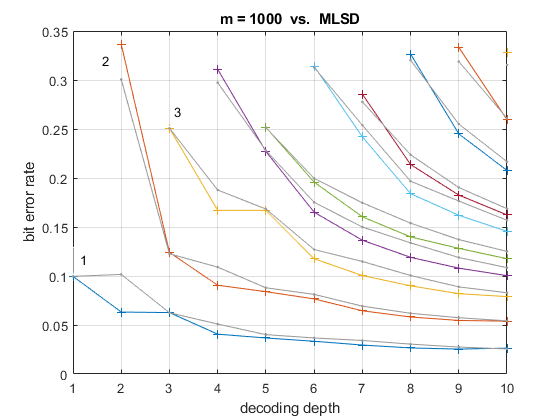}
	\caption{BER for multiple rounds of MCTS decoding supporting anytime reliability, with $d=10$ and $m=1000$. Grey curves indicate BER for MLSD.}
	\label{fig:d10_6_c10_p1_m1000_n120000_hi_vs_ml}
\end{figure}

Figures~\ref{fig:d10_6_c10_p1_m10_n120000_hi} to~\ref{fig:d10_6_c10_p1_m1000_n120000_hi_vs_ml} show the bit error rate (BER) for multiple rounds of MCTS decoding that supports anytime reliability, with the code with depth $10$. The number of rounds of search in each round of decoding is set to $m=10$, $100$, and $1000$, respectively. The grey curves in Figure~\ref{fig:d10_6_c10_p1_m100_n120000_hi_vs_ml} and Figure~\ref{fig:d10_6_c10_p1_m1000_n120000_hi_vs_ml} indicate BER for MLSD. From these figures it is seen that 1) the decoding performance increases as $m$ is increased: when $m$ is increased to $1000$, the BER of MCTS decoding is even lower than the MLSD, which should not be surprising, as MLSD minimizes the sequence error rate, not necessarily the BER; and 2) for each bit, the BER decreases as the decoding delay increases, namely, when it is decoded in a later round.
We can also see the unequal error protection ability of this coding and decoding scheme, that is, the bits sent earlier are better protected against the channel noise, while the most recently sent bits can have a decoding error rate even larger than the channel transition probability. This is a natural property of random tree codes. By increasing the code length, more bits can be protected by the increased search depth.

\begin{figure}[h]
	\centering
	\includegraphics[scale = 0.62]{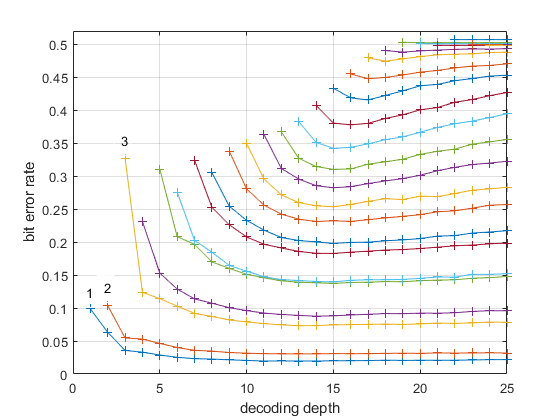}
	\caption{BER for multiple rounds of MCTS decoding supporting anytime reliability, with $d=25$ and $m=1000$.}
	\label{fig:d25_4_c25_p1_m1000_n10451_hi_ber}
\end{figure}

\begin{figure}[h]
	\centering
	\includegraphics[scale = 0.62]{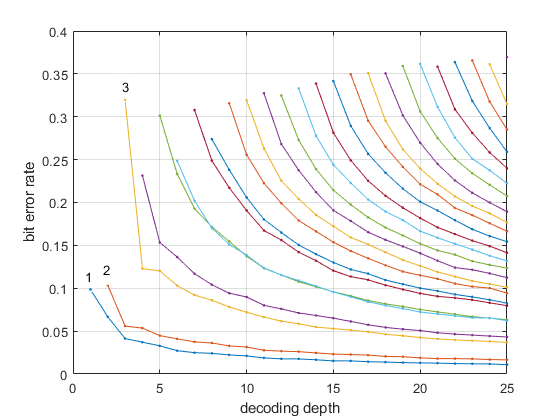}
	\caption{BER for MLSD, with $d=25$.}
	\label{fig:d25_4_c25_p1_m1000_n10451_ml_ber}
\end{figure}

Figures~\ref{fig:d25_4_c25_p1_m1000_n10451_hi_ber} and~\ref{fig:d25_4_c25_p1_m1000_n10451_ml_ber} show the same BER plots for the code with depth $25$. One is for the MCTS decoding with $m=1000$, and the other is for MLSD. 
It can be seen that with a much lower computation complexity, the MCTS decoding has comparable performance with MLSD for bits with indices $i\le 4$. At the same time, the BER for the bits with large index can be much higher than those with smaller indices, which is a downside to have all the bits decoded in a single-round MCTS decoding. To further improve the decoding performance, one could either increase $m$, or perform sliding-root MCTS decoding.

\begin{figure}[h]
	\centering
	\includegraphics[scale = 0.62]{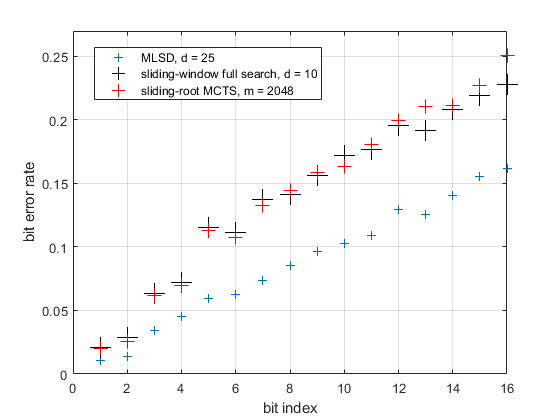}
	\caption{BER for sliding-root MCTS decoding, with $d=25$ and $m=2048$, compared with BER for sliding-window full search decoding with search depth $10$ and BER for MLSD.}
	\label{fig:d25_4_c25_p1_m2048_n4599_ml_fs10}
\end{figure}

Figure~\ref{fig:d25_4_c25_p1_m2048_n4599_ml_fs10} compares the BERs for the sliding-root MCTS decoding, the sliding-window full search decoding, and the MLSD, on the code with depth $25$. For the sliding-root MCTS decoding, every search is run to the leaf node and $m=2048$ rounds of search is run in each round of decoding; while for the sliding-window full search decoding, the search depth is set to $10$, so that these two decoding methods have roughly the same computational complexity. It can be seen that for bits with smaller indices, e.g. $i\le 7$, the sliding-root MCTS decoding performs slightly better than the sliding-window full search decoding under similar computation complexity.

%
%

\section{Possible improvements and extensions}
Some possible improvements and extensions of the proposed decoding method are listed below.
\begin{itemize}[leftmargin=*]
	\item
	\emph{Reuse of $Q(s,a)$}
	
	When multiple rounds of MCTS decoding are performed, the values of $Q(s,a)$ obtained in each round of decoding may be reused for the future rounds.
	This reuse may improve the decoding performance without a heavy increase of the computation complexity.	
	Following the practice in AlphaGo \cite{AlphaGo,AlphaGoZero}, $Q(s,a)$ as a function of $(s,a)$ can even be represented by neural networks and updated by a reinforcement learning scheme.

	\item
	\emph{Support for soft output and/or soft input}
	
	The proposed decoding method can be modified to support the soft output, or reliability, for each information symbol.
	For example, with the sliding-root MCTS decoding, the probability of the $i$th information symbol being $a$ can be computed as $\frac{1}{Z}e^{\beta Q(s,a)}$, where $s$ is the node determined by the hard decisions of the previous information symbols, $\beta$ is a parameter that can be tuned, and $Z$ is a normalization factor.
	The soft input may also be supported by incorporating any prior knowledge into the initialization functions $Q_0$ and $N_0$.

	\item
	\emph{Decoding by Monte-Carlo trellis search}
	
	The proposed decoding method based on MCTS can be naturally extended to decoding codes with a trellis structure. Instead of searching over a tree, the proposed method can be used for searching for the minimum-length path over a trellis, which amounts to the MLSD decoding of the trellis codes over DMC. The decoding method may be termed as decoding by Monte-Carlo \emph{trellis} search.

	\smallskip
	\item
	\emph{Decoding irregular tree codes}
	
	The proposed decoding method can also be used for irregular tree codes, e.g. the recently proposed polar-adjusted convolutional (PAC) codes \cite{pac_code}.
\end{itemize}

	\bibliography{aolin.bib}

\begin{thebibliography}{10}
\providecommand{\url}[1]{#1}
\csname url@samestyle\endcsname
\providecommand{\newblock}{\relax}
\providecommand{\bibinfo}[2]{#2}
\providecommand{\BIBentrySTDinterwordspacing}{\spaceskip=0pt\relax}
\providecommand{\BIBentryALTinterwordstretchfactor}{4}
\providecommand{\BIBentryALTinterwordspacing}{\spaceskip=\fontdimen2\font plus
\BIBentryALTinterwordstretchfactor\fontdimen3\font minus
  \fontdimen4\font\relax}
\providecommand{\BIBforeignlanguage}[2]{{%
\expandafter\ifx\csname l@#1\endcsname\relax
\typeout{** WARNING: IEEEtran.bst: No hyphenation pattern has been}%
\typeout{** loaded for the language `#1'. Using the pattern for}%
\typeout{** the default language instead.}%
\else
\language=\csname l@#1\endcsname
\fi
#2}}
\providecommand{\BIBdecl}{\relax}
\BIBdecl

\bibitem{Wozencraft1957Sequential}
J.~Wozencraft, ``Sequential decoding for reliable communication,'' Research Lab
  of Electronics, MIT, Tech. Rep., 1957.

\bibitem{BCJR}
L.~{Bahl}, J.~{Cocke}, F.~{Jelinek}, and J.~{Raviv}, ``Optimal decoding of
  linear codes for minimizing symbol error rate,'' \emph{IEEE Transactions on
  Information Theory}, vol.~20, no.~2, pp. 284--287, 1974.

\bibitem{trellis_blk}
F.~R. {Kschischang} and V.~{Sorokine}, ``On the trellis structure of block
  codes,'' \emph{IEEE Transactions on Information Theory}, vol.~41, no.~6, pp.
  1924--1937, 1995.

\bibitem{Sahai_Mitter_06}
A.~Sahai and S.~Mitter, ``The necessity and sufficiency of anytime capacity for
  stabilization of a linear system over a noisy communication link, {P}art {I}:
  Scalar systems,'' \emph{IEEE Trans. Inform. Theory}, vol.~52, no.~8, pp.
  3369--3395, 2006.

\bibitem{Schulman96_iter_comm}
L.~Schulman, ``Coding for interactive communication,'' \emph{IEEE Trans.
  Inform. Theory}, vol.~42, no.~6, pp. 1745--1756, 1996.

\bibitem{Elias_conv}
P.~Elias, ``Coding for noisy channels,'' \emph{IRE Conv. Rec.}, pp. 37--46,
  Mar. 1955.

\bibitem{LTI_at_codes16}
R.~T. {Sukhavasi} and B.~{Hassibi}, ``Linear time-invariant anytime codes for
  control over noisy channels,'' \emph{IEEE Transactions on Automatic Control},
  vol.~61, no.~12, pp. 3826--3841, 2016.

\bibitem{at_ldpccc}
L.~{Grosjean}, L.~K. {Rasmussen}, R.~{Thobaben}, and M.~{Skoglund},
  ``Systematic {LDPC} convolutional codes: Asymptotic and finite-length anytime
  properties,'' \emph{IEEE Transactions on Communications}, vol.~62, no.~12,
  pp. 4165--4183, 2014.

\bibitem{at_spcc}
M.~{Noor-A-Rahim}, K.~D. {Nguyen}, and G.~{Lechner}, ``Anytime reliability of
  spatially coupled codes,'' \emph{IEEE Transactions on Communications},
  vol.~63, no.~4, pp. 1069--1080, 2015.

\bibitem{ECC_book_ShuLin}
S.~Lin and D.~J. Costello, \emph{Error Control Coding}, 2nd~ed.\hskip 1em plus
  0.5em minus 0.4em\relax Pearson, 2005.

\bibitem{digit_comm_book5}
J.~G. Proakis and M.~Salehi, \emph{Digital Communications}, 5th~ed.\hskip 1em
  plus 0.5em minus 0.4em\relax McGraw-Hill Education, 2007.

\bibitem{Remi_mcts06}
R.~Coulom, ``Efficient selectivity and backup operators in {Monte-Carlo} tree
  search,'' in \emph{Proceedings Computers and Games}.\hskip 1em plus 0.5em
  minus 0.4em\relax Springer-Verlag, 2006.

\bibitem{Kocsis06banditbased}
L.~Kocsis and C.~Szepesvári, ``Bandit based {Monte-Carlo} planning,'' in
  \emph{European Conference on Machine Learning}.\hskip 1em plus 0.5em minus
  0.4em\relax Springer, 2006.

\bibitem{AlphaGo}
D.~{Silver}, A.~{Huang}, C.~J. {Maddison}, A.~{Guez}, L.~{Sifre}, G.~{van den
  Driessche}, J.~{Schrittwieser}, I.~{Antonoglou}, V.~{Panneershelvam},
  M.~{Lanctot}, S.~{Dieleman}, D.~{Grewe}, J.~{Nham}, N.~{Kalchbrenner},
  I.~{Sutskever}, T.~{Lillicrap}, M.~{Leach}, K.~{Kavukcuoglu}, T.~{Graepel},
  and D.~{Hassabis}, ``{Mastering the game of Go with deep neural networks and
  tree search},'' \emph{Nature}, vol. 529, no. 7587, pp. 484--489, Jan. 2016.

\bibitem{AlphaGoZero}
D.~{Silver}, J.~{Schrittwieser}, K.~{Simonyan}, I.~{Antonoglou}, A.~{Huang},
  A.~{Guez}, T.~{Hubert}, L.~{Baker}, M.~{Lai}, A.~{Bolton}, Y.~{Chen},
  T.~{Lillicrap}, F.~{Hui}, L.~{Sifre}, G.~{van den Driessche}, T.~{Graepel},
  and D.~{Hassabis}, ``{Mastering the game of Go without human knowledge},''
  \emph{Nature}, vol. 550, no. 7676, pp. 354--359, Oct. 2017.

\bibitem{Zilberstein_1996}
S.~Zilberstein, ``Using anytime algorithms in intelligent systems,'' \emph{AI
  Magazine}, vol.~17, no.~3, p.~73, Mar. 1996.

\bibitem{adp_samp_mdp}
H.~S. Chang, M.~C. Fu, J.~Hu, and S.~I. Marcus, ``An adaptive sampling
  algorithm for solving {M}arkov decision processes,'' \emph{Operations
  Research}, vol.~53, no.~1, pp. 126--139, 2005.

\bibitem{dmuu05}
M.~J. Kochenderfer, \emph{Decision Making Under Uncertainty Theory and
  Application}.\hskip 1em plus 0.5em minus 0.4em\relax MIT Press, 2015.

\bibitem{pac_code}
E.~Arıkan, ``From sequential decoding to channel polarization and back
  again,'' \emph{IEEE Information Theorey Society Newsletter}, Sep. 2019.

\end{thebibliography}

\end{document}